\documentclass[conference]{IEEEtran}
\IEEEoverridecommandlockouts
\usepackage{cite}
\usepackage{amsmath,amssymb,amsfonts}
\usepackage{algorithmic}
\usepackage{graphicx}
\usepackage{hyperref}
\usepackage{textcomp}
\usepackage{xcolor}

\usepackage{tikz}
\usepackage{lipsum}

\def\BibTeX{{\rm B\kern-.05em{\sc i\kern-.025em b}\kern-.08em
    T\kern-.1667em\lower.7ex\hbox{E}\kern-.125emX}}

\newcommand\copyrighttext{%
  \footnotesize \textcopyright 2023 IEEE. Personal use of this material is permitted.
  Permission from IEEE must be obtained for all other uses, in any current or future 
  media, including reprinting/republishing this material for advertising or promotional 
  purposes, creating new collective works, for resale or redistribution to servers or 
  lists, or reuse of any copyrighted component of this work in other works. 
 }
\newcommand\copyrightnotice{%
\begin{tikzpicture}[remember picture,overlay]
\node[anchor=south,yshift=10pt] at (current page.south) {\fbox{\parbox{\dimexpr\textwidth-\fboxsep-\fboxrule\relax}{\copyrighttext}}};
\end{tikzpicture}%
}

\begin{document}

\newpage
\title{Reconstructing Atmospheric Parameters of Exoplanets Using Deep Learning \\

}

\newcommand{\todo}[1]{{\color{red}[#1]}}

\author{
\IEEEauthorblockN{Flavio Giobergia}
\IEEEauthorblockA{
\textit{Politecnico di Torino}\\
Turin, Italy \\
flavio.giobergia@polito.it}
\and
\IEEEauthorblockN{Alkis Koudounas}
\IEEEauthorblockA{
\textit{Politecnico di Torino}\\
Turin, Italy \\
alkis.koudounas@polito.it}
\and
\IEEEauthorblockN{Elena Baralis}
\IEEEauthorblockA{
\textit{Politecnico di Torino}\\
Turin, Italy \\
elena.baralis@polito.it}
}


\maketitle
\copyrightnotice


\begin{abstract}
Exploring exoplanets has transformed our understanding of the universe by revealing many planetary systems that defy our current understanding. 
To study their atmospheres, spectroscopic observations are used to infer essential atmospheric properties that are not directly measurable. 
Estimating atmospheric parameters that best fit the observed spectrum within a specified atmospheric model is a complex problem that is difficult to model.
In this paper, we present 
a multi-target probabilistic regression approach that combines deep learning and inverse modeling techniques within a multimodal architecture to extract atmospheric parameters from exoplanets.
Our methodology overcomes computational limitations and outperforms previous approaches, enabling efficient analysis of exoplanetary atmospheres. This research contributes to advancements in the field of exoplanet research and offers valuable insights for future studies.
\end{abstract}


\section{Introduction}
Exploring exoplanets has revolutionized our knowledge of the universe, unveiling a diverse range of planetary systems that challenge our existing notions~\cite{batalha2014exploring}. 
These exoplanets span a wide spectrum, ranging from Earth-like habitable worlds~\cite{kaltenegger2017characterize} to searing hot-Jupiters~\cite{dawson2018origins} and everything in between. Their existence offers profound insights into the mechanisms governing planetary formation and evolution.

To deepen our understanding and contextualize our own solar system within the broader galactic landscape, the European Space Agency (ESA) is collaborating with NASA and JAXA to develop the Ariel Space Mission~\cite{tinetti2016science}. 
This ambitious endeavor aims to extensively investigate the atmospheres of hundreds of exoplanets in close proximity using a meter-class telescope in L2. By analyzing spectra and photometric data in the visible and infrared range, Ariel will extract chemical compositions and study the thermal properties of exoplanetary atmospheres. The mission's open data policy will allow rapid access to high-quality exoplanet spectra for scientific research.

While the Ariel mission promises groundbreaking discoveries, it faces a formidable hurdle in the complex modeling of exoplanetary atmospheres. These atmospheres exhibit intricate chemistries, cloud formations, and dynamic processes, making their characterization a highly intricate task. The traditional approach of forward modeling, which involves simulating atmospheric spectra based on predefined models, is computationally intensive and unsuitable for efficient fitting or sampling techniques such as Markov Chain Monte Carlo or Nested Sampling~\cite{Madhusudhan2018}.

This paper presents our proposed solution to the Ariel Machine Learning Data Challenge\footnote{\url{https://www.ariel-datachallenge.space/}}, which involves addressing the multi-target probabilistic regression problem of extracting critical atmospheric parameters of exoplanets on the Ariel Big Challenge (ABC) dataset~\cite{changeat2023esa}. 
We propose an innovative multimodal architecture that leverages deep learning and inverse modeling techniques to map observed atmospheric signatures to the underlying planetary characteristics. 
In addition, we incorporate a modeling approach where the target output is treated as a Gaussian distribution with a full covariance matrix. This enables us to predict and estimate the parameters $\mu$ (mean) and $\Sigma$ (covariance matrix) that define the distribution.
By employing a data-driven approach, our method aims to overcome the limitations of conventional forward modeling and expedite the characterization process, enabling rapid analysis of exoplanetary atmospheres. 

Our contributions are as follows:
\begin{itemize}
    \item We introduce a novel deep learning multimodal approach that addresses the multi-regression problem of determining key atmospheric parameters of exoplanets, outperforming previous approaches. 
    \item We overcome the computational limitations of traditional forward modeling techniques by leveraging data-driven methodologies. Our approach allows for faster and more efficient analysis of exoplanetary atmospheres.
    \item Our work contributes to the broader field of exoplanet research by providing a more efficient and accurate method for characterizing exoplanetary atmospheres. 
\end{itemize}

The insights gained from our approach have practical implications for future space missions, including the Ariel Space Mission. 
Our methodology, which achieved the $8^{th}$ position in the highly competitive competition with over 250 participants, provides a promising approach to accelerate the analysis of exoplanetary atmospheres. 
This approach enables quicker and more thorough characterization of numerous exoplanets, contributing to advancements in the field.

The remainder of the paper is organized as follows. We first examine the related works and highlight the difference between the present work and existing approaches (Section~\ref{sec:rw}). We then detail the problem we have to address (Section~\ref{sec:problem}) and the dataset utilized for the experiments, and we discuss the methodology employed in our approach (Section~\ref{sec:methodology}). In Section~\ref{sec:results}, we present our experimental results and finally highlight the implications and potential applications of our findings (Section~\ref{sec:conclusion}). 

\section{Related Works}\label{sec:rw}

The field of exoplanet studies has evolved from detecting exoplanets to characterizing their atmospheres through retrieval techniques. Retrieval is an inverse modeling approach that compares forward models of a planet's spectrum to observational data, allowing for the estimation of atmospheric parameters~\cite{deming2017illusion, himes2022accurate}. 
Atmospheric retrieval is thus crucial in helping astronomers understand and analyze individual observations acquired through spectroscopy during transit, eclipse, and phase curves. 
This retrieval technique applies to both low and high-resolution data~\cite{mikal2022diurnal, mansfield2022confirmation, changeat2022spectroscopic, boucher2021characterizing, molliere2020retrieving}, making it an indispensable tool for astronomers in interpreting the atmospheric properties of exoplanets.

Bayesian methods are commonly used to retrieve noisy exoplanet spectra, providing posterior distributions that constrain model parameters and determine their statistical significance, leveraging sampling techniques such as Markov Chain Monte Carlo (MCMC) or Nested Sampling~\cite{Madhusudhan2018}. 
However, traditional Bayesian retrieval methods can be computationally expensive and time-consuming~\cite{zhang2018advances}.
As the volume of exoplanetary data increases, driven by missions like the James Webb Space Telescope (JWST)~\cite{gardner2006james} and Ariel~\cite{tinetti2016science}, alternative approaches to computing posterior distributions are needed.

Machine learning (ML) and deep learning (DL) techniques have been applied to many areas within exoplanetary science, including data detrending~\cite{morvan2021pylightcurve, krick2020random}, debris removal~\cite{lawler2012debris, koudounas2022time}, and planet detection and characterization~\cite{martinez2022convolutional, yip2020compatibility, yip2021peeking, yip2022sample}. 
Various ML and DL approaches, such as random forests, generative adversarial networks (GANs), convolutional neural networks (CNNs), and Bayesian neural networks, have been employed to improve retrieval efficiency while significantly reducing computational time. 
Although these approaches show promise, as computational time is reduced, the accuracy of the posterior estimation decreases due to substantial approximations made during Bayesian inference.

In contrast to previous approaches, our methodology incorporates a multimodal 1D-CNN that combines information from both the spectral data and auxiliary data related to stellar and planetary parameters. This novel approach allows us to achieve favorable results in approximating the posterior distribution while maintaining remarkably low computational time.

\section{Problem Statement}\label{sec:problem}
Exoplanets are discovered using various methods, but the most commonly used techniques are radial velocity and transit~\cite{changeat2023esa}. Transit is an indirect method to monitor changes in the brightness of the host star. During a transit event, the planet passes in front of the star, causing a reduction in the amount of light observed from Earth. By studying these transit events at different wavelengths, astronomers can gain insights into the atmospheric properties of the exoplanet.

The exoplanetary atmosphere affects the transit depth by absorbing stellar light wavelength-dependently. This absorption profile is influenced by the composition (molecular species, clouds, hazes) and characteristics (thermal structure) of the atmosphere. Astronomers employ simplified models to analyze the observed signal, known as a \textit{spectrum}, and investigate the underlying atmospheric processes occurring in exoplanet atmospheres.

The study of exoplanetary atmospheres relies on spectroscopic observations to infer fundamental atmospheric properties that cannot be directly measured. This process is known as the inverse problem~\cite{potthast2006survey}, where one tries to deduce the atmospheric properties based on the observed effects. However, the observed effects are often corrupted due to factors such as noise and limited spectroscopic coverage, leading to a loss of information, which can, in turn, result in model degeneracy.

The main objective of atmospheric retrieval is to estimate the parameters that best explain the observed spectrum under a given atmospheric model. This is typically approached using a forward model, which incorporates assumptions about the atmosphere and an optimizer. 
In the Bayesian framework, the goal is to determine the posterior distribution, which represents the conditional distribution of the model parameters given the observed data.

In our case, the aim of the Ariel Machine Learning Data Challenge is to predict the conditional joint distribution (or the Bayesian posterior distribution) of 7 atmospheric properties given the observed spectrum, namely planet radius (in Jupyter radii), temperature (in Kelvin), and the log-abundance of five atmospheric gases: $H_2O$, $CO_2$, $CO$, $CH_4$, and $NH_3$.
The probability densities of these variables ($\theta = \theta_1, \theta_2, ..., \theta_n$) given the observed data ($D$) are collectively referred to as the posterior distribution, $P(\theta | D)$.

\subsection{Dataset}
\label{ssec:data}

To develop and validate our approach, we utilize the Ariel Big Challenge (ABC) Database~\cite{changeat2023esa}, encompassing simulated atmospheric spectra, ground-truth models, and Bayesian posterior distributions for a total of 6,766 exoplanets. The database is generated using the powerful Alfnoor~\cite{mugnai2021alfnoor} framework combining the TauREx 3~\cite{al2021taurex} atmospheric modeling suite and the ArielRad~\cite{mugnai2020arielrad} instrument simulator.
The dataset comprises two types of data, namely spectral and auxiliary data. Additionally, the target data is available in the form of tracedata.

\begin{figure}
    \centering
    \includegraphics[width=\linewidth]{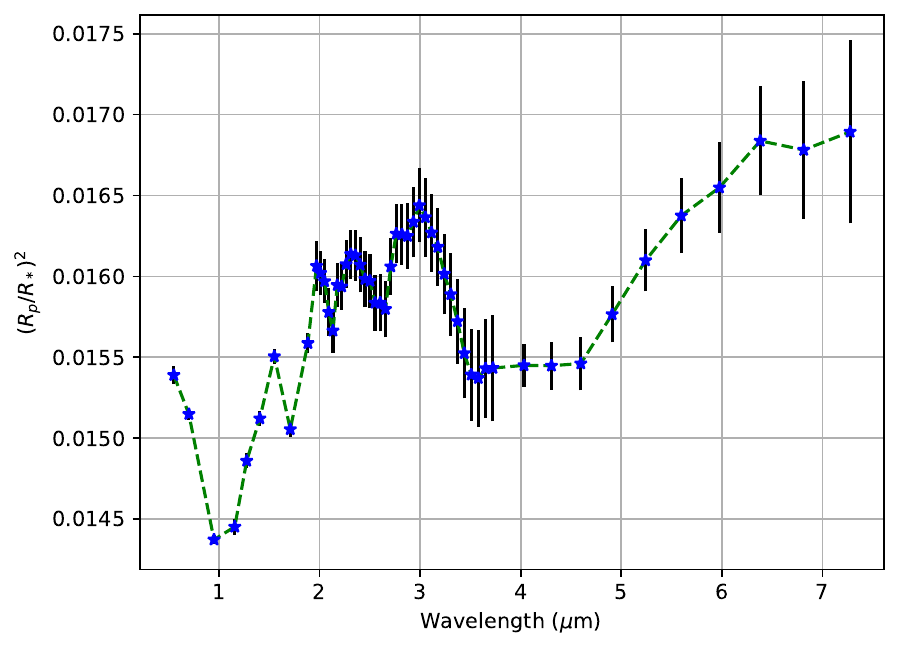}
    \caption{Spectral information available for each exoplanet. The quantity $(R_p/R_*)^2$ represents the ratio between the planet's radius, and that of the orbited star since the measured dip can be quantified in terms of a ratio of the areas of the two entities involved.}
    \label{fig:spectra}
\end{figure}

\vspace{2mm}
\noindent \textit{Spectral data.} Each entry in the dataset consists of an atmospheric spectrum with 52 data points. Each of them contains information about the intensity measure (transit depth), the corresponding wavelength of light, the spectral resolution (wavelength bin size), and the associated measurement uncertainty. Figure \ref{fig:spectra} represents the spectrum for one of the exoplanets available in the training set.

\vspace{2mm}
\noindent \textit{Auxiliary data.} Each example further encompasses eight additional stellar and planetary parameters as auxiliary information, including star distance, stellar mass, stellar radius, stellar temperature, planet mass, orbital period, semi-major axis, planet radius, and surface gravity. This information is unique to each planet and is sourced from various exoplanet datasets.

\begin{figure}
    \centering
    \includegraphics[width=\linewidth]{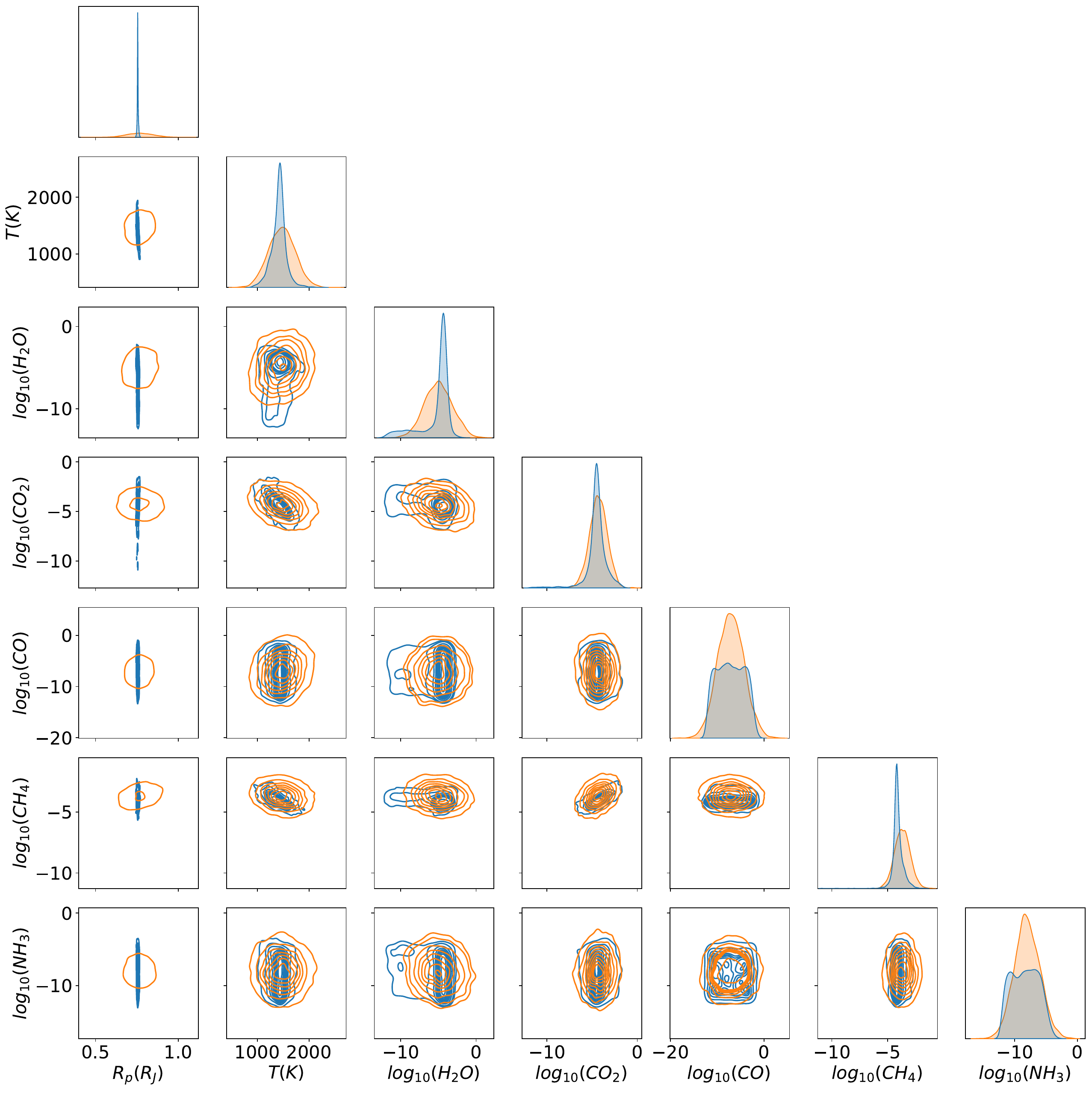}
    \caption{Samples for the atmospheric parameters for one of the available exoplanets. In blue is the ground truth distribution, in orange the distribution predicted by the proposed approach. }
    \label{fig:target-example}
\end{figure}

\vspace{2mm}
\noindent \textit{Tracedata.} The target output is provided in terms of samples from the conditional joint distribution, characterized by the aforementioned 7 atmospheric properties, referred to as tracedata. For each planet, an average of 3,938 samples are available (standard deviation of 654). Figure \ref{fig:target-example} shows one such distribution as an example.

\section{Methodology}\label{sec:methodology}
\begin{figure*}
    \centering
    \includegraphics[width=\linewidth]{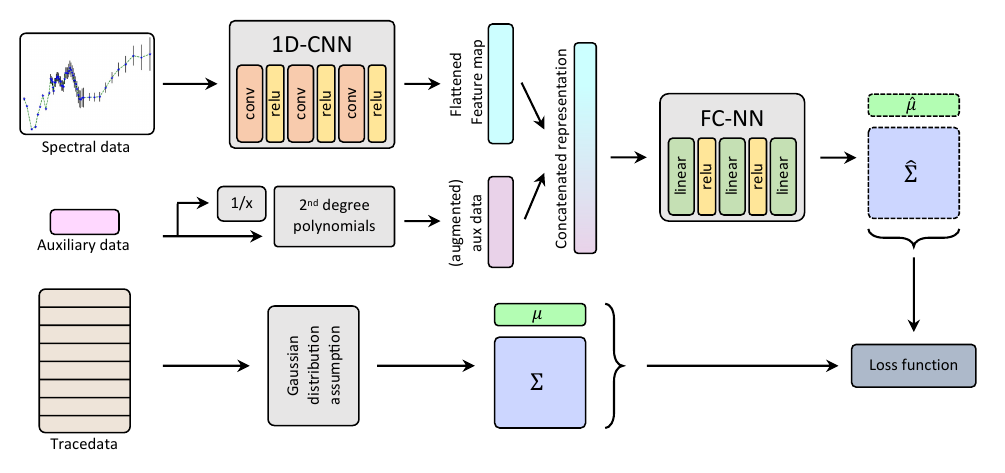}
    \caption{Architecture for the proposed solution. The spectral data is processed through a convolutional model (1D-CNN). The auxiliary data is augmented and merged with the processed spectral data. A final fully-connected model (FC-NN) maps the processed input to the desired probability distribution. The parameters for the target distribution are also produced and used when computing the loss function.}
    \label{fig:architecture}
\end{figure*}

We propose adopting a deep learning solution to solve the Ariel Machine Learning Data Challenge problem. At a high level, we leverage both the spectral data and the auxiliary data that has been made available with the adoption of a multimodal architecture; we additionally model the target output as a full covariance Gaussian distribution, thus making it possible to predict the parameters $\mu$ and $\Sigma$ that characterize it. Figure \ref{fig:architecture} summarizes the proposed approach. The rest of this section analyzes each component in further detail.

\subsection{Spectral data}
As already argued in Subsection \ref{ssec:data}, the spectral data conveys significant information regarding the composition of the atmosphere of an exoplanet. Since the atmospheric spectrum is continuous in nature, we consider adopting a 1-dimensional convolutional (1D-CNN) model to process the spectrum to be able to characterize local behaviors that can be found in the spectrum itself. The adoption of 1D-CNN models has already been adopted in literature with promising results \cite{yip2021peeking}. We build a representation of the spectrogram as a dense vector by flattening the feature maps obtained as the output of the 1D-CNN.

\subsection{Auxiliary data}
The auxiliary data available provides information on the planetary system, including the exoplanet under study and the orbited star. Based on the intuition that some of the auxiliary information may have non-linear relationships with the target outputs, we introduce a non-linear transformation of these inputs. In particular, we additionally obtain the inverse of the auxiliary quantities and, from these, we extract the polynomials up to degree 2. Starting from the original 8 features, we obtain a total of 152 transformed auxiliary features. The extracted features are then concatenated to the dense vector that is used to represent the spectral data. In this way, we produce a vector that represents the entire input both in terms of atmospheric spectrum and contextual planetary system information.

\subsection{Parameters estimation}
The main goal of this work is to produce samples that are drawn from an underlying, unknown data distribution. These samples represent different atmospheric models that make different assumptions (e.g., in terms of atmospheric composition). We frame the problem by modeling the underlying distribution to produce the desired samples as a byproduct. 

The target available is provided as samples from the ground truth distribution. Since we aim to predict this distribution, we need to make an assumption regarding the type of probability distribution. We assume the distribution to be a multivariate Gaussian one with full covariance: based on the intuition that can be extracted from Figure \ref{fig:target-example}, this is a simplifying assumption. However, it allows for a simple parametrization of the distribution. Additionally, it allows for the modeling of interactions across dimensions thanks to the adoption of a full covariance. 

For each planet, we focus on $n$ target variables $\theta_1, \theta_2, ..., \theta_n$ (for this specific challenge -- as discussed -- we have $n = 7$, but the same approach can be adopted for differently framed problems). We can defined the mean vector $\mu \in \mathbb{R}^n$ such that $\mu_i = E[T_i]$ and the covariance matrix $\Sigma \in \mathbb{R}^{n \times n}$ such that $\Sigma_{ij} = cov(T_i, T_j)$ and $\Sigma_{ii} = Var(T_i)$.

Since the covariance matrix is symmetric, we can build a compressed representation using $\frac{n(n+1)}{2}$ values. When accounting for $\mu$, the overall distribution can be parametrized with a total of $\frac{n(n+3)}{2}$ values. We refer to the target representation for the $k^{th}$ planet as $y^{(k)} \in \mathbb{R}^{\frac{n(n+3)}{2}}$. We note that we can freely move from the $\mu^{(k)}, \Sigma^{(k)}$ representation to the $y^{(k)}$ one. Similarly, we refer to the predicted parameters as either $\hat{y}^{(k)}$ or as the corresponding $\hat{\mu}^{(k)}, \hat{\Sigma}^{(k)}$.

\subsection{Loss function}
To compute the proximity of the predicted distribution with the target one, we considered various possible loss functions. We empirically observed that computing the KL divergence leads to a difficult convergence process. In contrast, we obtain better convergence properties by minimizing the L1 loss function between the predicted and ground truth values. In particular, the loss is defined as:

\begin{equation}
    \ell(y, \hat{y}) = | y - \hat{y} |
\end{equation}

This is equivalent, in terms of mean vectors and covariance matrices, as:

\begin{multline}
    \ell(\mu_i, \Sigma_i, \hat{\mu}_i, \hat{\Sigma}_i) = 
    \sum_{0 \le i < n} \sum_{i < j < n} | \Sigma_{ij} - \hat{\Sigma}_{ij} | + \\
    \sum_{0 \le i < n} | \Sigma_{ii} - \hat{\Sigma}_{ii} | +
    \sum_{0 \le i < n} | \mu_i - \hat{\mu}_i | 
\end{multline}

\section{Results}\label{sec:results}
In accordance with the policies of the Ariel Machine Learning Data Challenge, we evaluate the quality of the obtained results using two terms: the posterior score and the spectral score. 
The two metrics complement each other: the first ensures that the predictions accurately replicate each individual distribution for the variables involved, while the second focuses on preserving the physical laws or, more precisely, the interrelationships (covariance) among different targets. 

We employ the 2 Sample Kolmogorov–Smirnov test (K-S test) to evaluate the posterior score scenario. The K-S test is a widely used statistical test determining whether two given samples are derived from the same continuous distribution underlying them. The output is adjusted to range from 0 to 1000, where a score of 1000 indicates the highest similarity.

Regarding the spectral score, we compare the ``median'' values in spectral space, which consists of the median value for each wavelength bin, along with their uncertainty bounds (interquartile ranges of each wavelength bin), against the corresponding values obtained from Bayesian Nested Sampling. We quantify the differences using an inverse Huber loss. The spectral score is calculated as a linear combination of the differences between the bounds and the differences between the median spectra. The maximum score is set to 1000, with 0 representing the lowest similarity.

The final score is computed as a weighted sum of the spectral loss and posterior loss:
\begin{equation}
    Final = (1 - \beta) \cdot Spectral + \beta \cdot Posterior
\end{equation}
where $\beta$ is set to $0.8$. The value set for $\beta$ is in accordance with the one adopted for the Ariel Machine Learning Data Challenge and reflects the greater importance that is generally assigned to the posterior score w.r.t. the spectral one. The final score ranges from a minimum of 0 to a maximum of 1000.

Table \ref{tab:results} summarizes the results achieved by the baseline model presented in \cite{yip2021peeking} and by the proposed approach on a test set obtained as a 20\% hold-out on the available data. For completeness, we include the results obtained by the proposed methodology when changing the loss function to another commonly adopted one, the mean squared error (L2).

We show that the proposed approach consistently outperforms the baseline model in terms of overall (final) score. This is the result of a model that performs consistently well in terms of both posterior and spectral scores. By contrast, the method proposed in \cite{yip2021peeking} is unbalanced toward performing significantly well in terms of spectral score with a consistent drop in performance as far as the posterior score is concerned.

\begin{table}[htbp]
\caption{Results on the Ariel Big Challenge database in terms of posterior, spectral, and final score. Best results are highlighted in bold.}
\label{tab:results}
\begin{center}
\resizebox{\linewidth}{!}{
\begin{tabular}{|c|c|c|c|}
\hline
\textbf{Method} & \textbf{Posterior Score} & \textbf{Spectral Score} & \textbf{Final Score}
\\
\hline
Yip et al.~\cite{yip2021peeking} 
    & $331.58 \pm 60.57$ 
    & $\mathbf{880.05 \pm 4.75}$
    & $441.28 \pm 48.97$ \\
Ours 
    & $611.77 \pm 14.52$ 
    & $658.57 \pm 9.63$ 
    & $\mathbf{621.13 \pm 9.95}$ \\
Ours w/ L2 
    & $610.054 \pm 11.10$
    & $640.35 \pm 12.77$ 
    & $616.10 \pm 9.31$ \\
Ours w/ L1+L2 
    & $\mathbf{613.04 \pm 15.96}$ 
    & $637.07 \pm 11.26$ 
    & $617.71 \pm 15.33$ \\
\hline
\end{tabular}
}
\end{center}
\end{table}

\subsection{Qualitative results}
Figure \ref{fig:target-example} shows a ground truth distribution (in blue), along with the predicted one (in orange). We observe how there is a reasonably good fit for most variables, considering the constraints introduced by the Gaussian simplifying assumption. We can also notice how using a full covariance allows for better modeling of variables that show some degree of correlation.

However, we note that the proposed model particularly struggles with predicting the planet's radius. Indeed, the distribution of samples for that parameter is rather narrow. In contrast, the model makes a high-variance prediction: this produces a non-zero overlap with the target variable, but it is still far from correct or meaningful modeling.

\section{Conclusion and Future Works}\label{sec:conclusion}
In this paper, we presented a possible solution to identifying atmospheric parameters in exoplanets. The proposed solution leverages both spectral data and auxiliary information available on the planets to produce a meaningful estimate of the parameters of interest. We show that our methodology outperforms the baseline model of reference by producing a balanced prediction regarding both posterior and spectral scores, allowing us to reach $8^{th}$ place in the Ariel Machine Learning Challenge.

By qualitatively inspecting some of the predicted parameters, we observe how the model struggles to reconstruct some of them while performing rather well on others. We aim to improve the performance of the proposed solution by addressing this aspect and by exploring whether consistent patterns exist in the characteristics of the worse reconstructed exoplanets.

\section*{Acknowledgment}
This study was carried out within the FAIR - Future Artificial Intelligence Research and received funding from the European Union Next-GenerationEU (PIANO NAZIONALE DI RIPRESA E RESILIENZA (PNRR) – MISSIONE 4 COMPONENTE 2, INVESTIMENTO 1.3 – D.D. 1555 11/10/2022, PE00000013), as a part of the MALTO (MAchine Learning @ poliTO) team, with partial support by SmartData@PoliTO center on Big Data and Data Science. This manuscript reflects only the authors' views and opinions, neither the European Union nor the European Commission can be considered responsible for them. 

\bibliographystyle{IEEEtran}
\bibliography{conference_101719}

\end{document}